\documentclass{JCIN22}
\usepackage{algorithm}
\usepackage{algorithmic}
\usepackage{supertabular}
\usepackage{subfig}
\usepackage{gbt7714}

\begin{document}

\ArticleType{Research paper}
\Year{2022}

\title{CSGO: Generalized Optimization for Cold Start in Wireless Collaborative Edge LLM Systems}

\author{Xuran Liu}
\author{Nan Xue}
\author{Rui Bao}
\author{Yaping Sun}
\author{Zhiyong Chen}
\author{Meixia Tao}
\author{Xiaodong Xu}
\author{Shuguang Cui}

\maketitle

{\bf\textit{Abstract---} While deploying large language models on edge devices promises low-latency and privacy-preserving AI services, it is hindered by limited device resources. Although pipeline parallelism facilitates distributed inference, existing approaches often ignore the cold-start latency caused by on-demand model loading. In this paper, we propose a latency-aware scheduling framework that overlaps model loading with computation and communication to minimize total inference latency. Based on device and model parameters, the framework dynamically adjusts layer partitioning and allocation to effectively hide loading time, thereby eliminating as many idle periods as possible. We formulate the problem as a Mixed-Integer Non-Linear Program and design an efficient dynamic programming algorithm to optimize model partitioning and device assignment. Experimental results show that the proposed method significantly reduces cold-start latency compared to baseline strategies.\\[-1.5mm]

}

\barefootnote{X.R. Liu, N.Xue, R. Bao, Z.Y. Chen, M.X. Tao. Cooperative Medianet Innovation Center, Shanghai Jiao Tong University, Shanghai 200240, China. (e-mail: \{3232794836, nan.xue, 851756936, zhiyongchen, mxtao\}@sjtu.edu.cn).\\\indent
Y.P. Sun. Department of Broadband Communication, Pengcheng Laboratory, Shenzhen 518000, China and Future Network of Intelligent Institute (FNii), the Chinese University of Hong Kong (Shenzhen), Shenzhen 518172, China. (e-mail: sunyp@pcl.ac.cn).\\\indent
X.D. Xu. Beijing University of Posts and Telecommunications, Beijing 100876, China and Department of Broadband Communication, Pengcheng Laboratory, Shenzhen 518000, China. (e-mail: xuxiaodong@bupt.edu.cn).\\\indent
S.G. Cui. School of Science and Engineering (SSE) and Future Network of Intelligent Institute (FNii), the Chinese University of Hong Kong (Shenzhen), Shenzhen 518172, China and Department of Broadband Communication, Pengcheng Laboratory, Shenzhen 518000, China. (e-mail: shuguangcui@cuhk.edu.cn).
}


\section{INTRODUCTION}\label{sec:I}
Recent years have witnessed a paradigm shift in artificial intelligence (AI), driven by the advent of Large Language Models (LLMs) and generative diffusion models \cite{diffusion}. These models exhibit unprecedented capabilities, revolutionizing domains from natural language processing to creative content generation. As their influence extends across diverse aspects of digital lives, there is a growing demand to move beyond cloud-centric deployments and embed these powerful AI functionalities directly into personal computing, particularly onto mobile devices such as smartphones. On-device inference holds the promise of a new wave of applications, offering advantages such as  real-time responsiveness, enhanced user privacy through local data processing \cite{privacy}, and highly personalized user experiences.

\begin{figure}[!t]
  \centering
  \includegraphics[width=0.4\textwidth]{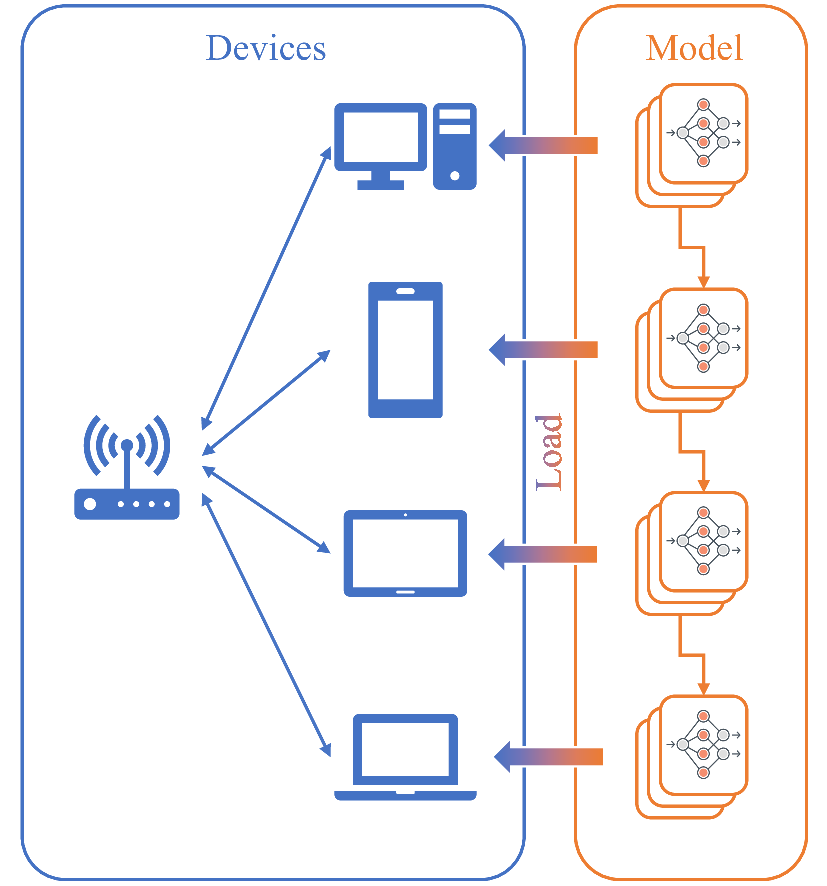}\\
 \caption{Pipeline-parallel model deployment on edge wireless networks with heterogeneous devices. The system adopts a star topology, where all devices communicate via a single access point (AP). The model is partitioned into sequential stages and distributed across devices for pipelined execution.}
  \label{fig:fig1}
\end{figure}

However, realizing this vision poses a fundamental challenge: the massive computational and memory requirements of state-of-the-art models sharply contrast with the limited resources of mobile devices. A single large model can easily exceed the available RAM and processing power of a typical smartphone, rendering local execution impractical. To address this mismatch, distributed inference has emerged as a promising paradigm \cite{inference}. The key idea is to partition a monolithic model into a smaller segments, or ``shards'', and distribute them across a network of collaborating mobile devices, as depicted in Fig.~\ref{fig:fig1}. By organizing these shards into a pipeline parallel workflow, the system can process inference requests in an assembly-line fashion, leveraging the aggregate capabilities of multiple devices to overcome individual hardware constraints.

\subsection{Related Work}
Pipeline Parallelism (PP)\cite{parallelization}, a fundamental model parallelism strategy, has been widely adopted for both the training and inference of Deep Neural Networks (DNNs).

In cloud-centric server environments, a variety of distributed frameworks have been developed to support large-scale model deployment, including GPipe, PipeDream, GraphPipe, ZeRO, and PipeFusion \cite{gpipe,pipedream,graphpipe,zero,pipefusion}. These systems partition models across multiple compute devices to overcome single-device memory limitations and accelerate computation. However, they are designed with the assumption of deployment in data centers, where abundant computational and memory resources, along with stable network connectivity, are readily available. In such settings, models are typically preloaded into memory and kept resident to serve a high volume of continuous requests--a model we refer to as ``always-on'' service.

This paradigm contrasts sharply with edge deployment scenarios, particularly on personal devices like smartphones or laptops, where memory, compute, and energy are highly constrained. To address these challenges, several lightweight pipeline parallel frameworks have emerged, such as PipeEdge, EdgePipe, EPipe, and WDMoE \cite{pipeedge,edgepipe,epipe,wdmoe}. These frameworks aim to optimize throughput for continuous inference tasks and are generally targeted at edge servers like smart cameras or IoT gateways, where models are loaded and executed over extended periods. However, they are less suited for personal edge devices, where inference requests tend to be sporadic and infrequent, and resources may not permit persistent model residency.

In addition, model compression methods such as model quantization\cite{hawq,awq}, pruning\cite{dep,playing}, and distillation\cite{minillm,adversarial} are also widely used to improve the performance of lightweight models suitable for deployment on edge devices.

However, these methods do not address the often-overlooked \textbf{cold start} problem. On personal devices, where long idle periods between inference requests are typical, each task may require loading the model from storage into memory from scratch. This results in substantial latency, which becomes a crucial bottleneck that directly impacts user experience. In this paper, we aim to address this overlooked yet crucial challenge.

\subsection{Contributions}
To address this cold-start challenge in wireless distributed collaborative inference, this paper proposes a \textbf{latency-aware pipeline scheduling algorithm} tailored for edge environments. The key idea is to overlap model loading with ongoing computation and communication phases within the pipeline. By strategically scheduling the on-demand loading of downstream shards in parallel with the execution of upstream stages, the proposed method effectively hides the loading latency, reduces pipeline stalls, and minimizes end-to-end inference latency.

The main contributions of this work are summarized as follows:
\begin{itemize}
    \item We systematically identify and analyze the cold-start latency issue in mobile distributed cooperative inference scenarios. Building upon a pipeline-parallel framework, we mathematically model the latency introduced by the cold start of large models.
    \item We design and implement a dynamic programming (DP)-based scheduling algorithm that jointly determines optimal layer partitioning and devices assignment for both loading and computation, with the goal of minimizing the end-to-end inference latency.
    \item We validate our approach through extensive experiments, demonstrating that the proposed method substantially reduces user-perceived latency and improves pipeline efficiency compared to conventional strategies that decouple loading from computation.
\end{itemize}

The rest of this paper is organized as follows. Section II presents the system model of collaborative edge LLMs inference. Section III formulates the optimization problem. Section IV proposes DP algorithm tailored for the latency optimization. Experimental results are shown in Section V and conclusions are drawn in Section VI.

\section{SYSTEM MODEL}\label{sec:II}
\subsection{Large Model Details}
We aim to deploy a large Transformer-based model\cite{transformer} onto a wireless network illustrated in Fig.~\ref{fig:fig1}. In the model, we focus exclusively on the Transformer Blocks, omitting the embedding and output layers. Let the model consists of $L$ Transformer Block layers, indexed by the set $\mathcal{L}\triangleq\{1,2,...,L\}$. In the edge deployment scenarios of interest, the number of model layers typically exceeds the number of available devices, i.e., $L\ge K$. Each layer $l\in\mathcal{L}$ is characterized by a triple $\phi_l = \{W_l,A_l,D_l\}$, where $W_l$ is the computational workload in floating point operations (FLOPs), $A_l$ is the size of the activations in megabytes (MB), $D_l$ is the parameter size in MB.

The model is a Transformer-based LLM that employs a multi-head attention mechanism with Grouped Query Attention (GQA)\cite{GQA} and a Swish-Gated Linear Unit (SwiGLU)\cite{GLU}-based feed-forward network (FFN). Both activations and parameters are represented using the bfloat16 (bf16) format.

The input representation has dimensionality $d_{\text{model}}$. The attention layer includes $h_q$ query heads and $h_k=h_v$ key-value heads, each with dimensionality $d_{\text{head}}$. The intermediate dimensionality of the feed-forward network is denoted by $d_{\text{ff}}$. To enable the calculation of concrete metrics such as computational cost and activation size, we define the input sequence length as $t$.

Given the above definitions, the metrics are calculated as follows:
    \begin{align}
        W_l^{\text{ATTN}} &= 4 t d_{\text{head}} (d_{\text{model}} h_q + d_{\text{model}} h_k + t h_q),\\
        W_l^{\text{FFN}} &= 6 t d_{\text{model}} d_{\text{ff}},\\
        W_l &= W_l^{\text{ATTN}} + W_l^{\text{FFN}}.
    \end{align}
Here,$W_l^{\text{ATTN}}$ includes the FLOPs associated with the query, key, and value (Q/K/V) projections, the context computation, and output projection. Similarly, $W_l^{\text{FFN}}$ accounts for the FLOPs from the up-projection, gate-projection, and down-projection within the FFN. Both additions and multiplications involved in matrix operations are included in the FLOP count. However, overheads from nonlinear activations functions and element-wise operations are omitted.

Similarly, based on the definitions of GQA and SwiGLU, the activation size and the parameter size of the model layer can be computed using the following formulas:
    \begin{align}
        A_l &= 2 t d_{\text{model}},\\
        P_l^{\text{ATTN}} &= 4 d_{\text{model}} d_{\text{head}} (h_q + h_k),\\
        P_l^{\text{FFN}} &=  6 d_{\text{model}} d_{\text{ff}},\\
        P_l &= P_l^{\text{ATTN}} + P_l^{\text{FFN}}.
    \end{align}
Here, $P_l^{\text{ATTN}}$ includes weights for the Q/K/V projections and the output projection. $P_l^{\text{FFN}}$ includes weights for the up-projection, gate-projection, and down-projection.

\subsection{System Resource Modeling}
We consider a wireless network consisting of a single AP and $K$ mobile devices, indexed by the set $\mathcal{K}\triangleq\{1,2,...,K\}$. These devices and the AP form a star topology, where communication between any tow devices occurs via the AP. Typical deployment scenarios include indoor Wi-Fi networks and cellular networks. Each device $k\in\mathcal{K}$ is characterized by a profile $\psi_k=\{c_k,b_k^{u},b_k^{d},r_k,m_k\}$, where $c_k$ is the computational capability in FLOPS, $b_k^{u}$ is the uplink bandwidth from mobile devices to the AP in Mbps, $b_k^{d}$ is the downlink bandwidth from the AP, measured in Mbps, $r_k$ is the disk read speed in MB/s, and $m_k$ is the GPU memory capacity in GB.

\subsubsection{Computational Resources}
We focus on low-workload inference scenarios, which are typical in edge deployments. In such setting, the actual utilization of a mobile device is often substantially lower than its theoretical peak performance or the average utilization observed during large-scale training\cite{programming}. Depending on factors such as batch size and input sequence length, utilization rates can vary widely, typically ranging from 1\% to 40\%. As a result, it is important to account for workload size when evaluating the effective computational capability.  

To capture this behavior, we model utilization using a saturating exponential function:
\begin{equation}
    U(t)=a (1-e^{-bt}),
\end{equation}
where $U(t)$ denotes the utilization rate as a function of the workload $t$, $a$ represents the maximum achievable utilization, and $b$ controls the growth rate. The parameters $a$ and $b$ are derived from empirical observations in preliminary experiments. Let $c_k^*$ denote the theoretical peak computational capability of device $k$. The effective computational capability, accounting for workload-dependent utilization, is then given by:
\begin{equation}
    c_k=c_k^*U(t).
\end{equation}

\subsubsection{I/O Resources}
During model cold start phase, the primary I/O bottleneck is the disk read speed\cite{computer}. As a result, we use the disk read speed as a proxy for the device's overall I/O performance. Since this metric is generally stable and the underlying-hardware-level read process lies outside the scope of our optimization, we abstract it using standardized, canonical values. The specific values adopted in our analysis are provided in Section V.

\subsubsection{Communication Resources}
Let $B_k$ denote the channel bandwidth allocated to device $k$, and $P_k^u$ and $P_k^d$ represent its uplink and downlink transmission power, respectively. Let $d_k$ be the distance from the device to the AP and $N_0$ be the noise power spectral density. Furthermore, to reflect practical constraints, an efficiency factor $\mu$ is applied to account for protocol overheads and practical limitations\cite{performance,survey}.

The achievable uplink and downlink data rates for device $k$ are formulated as follows\cite{shannon}:
\begin{align}
    b_k^{u}&=\mu B_k\log_2{(1+\frac{P_k^ug_k}{N_0B_k})},\\
    b_k^{d}&=\mu B_k\log_2{(1+\frac{P_k^dg_k}{N_0B_k})},
\end{align}
where $g_k$ represents the channel gain, which is modeled using a path loss model as follows\cite{wireless}:
\begin{equation}
    g_k=\beta_0(\frac{d_k}{d_0})^{-\zeta}
\end{equation}
where $\beta_0$ is the channel gain at a reference distance $d_0$, and $\zeta$ is the path loss exponent. For simplicity, we consider the channel gain $g_k$ is symmetric for both uplink and downlink transmissions.

\subsection{Deployment framework}\label{sec:II(C)}
This paper adopts a pipeline parallelism strategy to deploy the model across $K$ heterogeneous mobile devices, as depicted in Fig.2. The large model's $L$ layers are partitioned into $N$ contiguous, non-overlapping segments (hereafter referred to as partitions), where $1\le N\le K$. Each partition is assigned exclusively to a single device for execution.

To formalize the partitioning scheme, we define a triplet $\{k_n, s_n, e_n\}$ for each partition $n \in \mathcal{N} \triangleq \{1, 2, ..., N\}$. Here, $k_n \in \mathcal{K}$ denotes the index of the device assigned to the $n$-th partition, while $s_n \in \mathcal{L}$ and $e_n \in \mathcal{L}$ represent the starting and ending layer indices of the partition, respectively.

The allocation scheme must satisfy the following constraints. First, devices assigned to different partitions must be unique:
\begin{equation}
    \forall n, n'\in\mathcal{N},n\neq n' \Rightarrow k_{n}\neq k_{n'}.
\end{equation}

Second, the partitioning must be contiguous and comprehensive, covering all layers of the model:
\begin{align}
    s_1 = 1&,\quad e_N = L,\\
    \forall n\in\mathcal{N}&,\quad s_n \le e_n,\\
    \forall n\in\mathcal{N}\setminus\{1\}&,\quad s_{n} = e_{n-1} + 1.
\end{align}
    
Before the inference task begins, device $k_n$ stores the model parameters for its assigned layers (from $s_n$ to $e_n$) in its local storage. When an inference request arrives, it triggers a cold start process consisting of three main stages: model loading, activation transfer, and forward computation.

During pipeline execution, the start time $t^{\text{start}}_{k_n}$ for device $k_n$ depends on two factors: the completion of its own model loading and the computation finish time $t^{\text{finish}}_{k_{n-1}}$ of the preceding device $k_{n-1}$. Due to the limited I/O bandwidth of mobile devices, we consider that model loading and transmission are performed sequentially, that is, a device cannot begin communicating until it has completed loading the model. The model loading start time and the output transmission completion time of device $k_n$ are given by:
\begin{align}
    t^{\text{start}}_{k_n}&=\max{(t^{\text{load}}_{k_n},t^{\text{finish}}_{k_{n-1}})},\\
    t^{\text{finish}}_{k_{n}}&=t^{\text{start}}_{k_n}+t^{\text{comm}}_{k_n}+t^{\text{comp}}_{k_n}.
\end{align}

Here, $t^{\text{finish}}_{k_0}$ is defined as 0. The terms $t^{\text{load}}_{k_n}$, $t^{\text{comm}}_{k_n}$ and $t^{\text{comp}}_{k_n}$ represent the model loading latency, activation transfer latency, and computation latency of device $k_n$, respectively. These latencies are calculated as follows:
\begin{align}
    t^{\text{load}}_{k_n}&=\frac{\sum_{l=s_n}^{e_n}{P_l}}{r_{k_n}},\\
    t^{\text{comm}}_{k_n}&=\frac{A_{s_{n}-1}}{\min{(b^{u}_{k_{n-1}},b^{d}_{k_n})}},\\
    t^{\text{comp}}_{k_n}&=\frac{\sum_{l=s_n}^{e_n}{W_l}}{c_{k_n}}.
\end{align}

Additionally, to account for device memory constraints, we have:
\begin{equation}
    m_{k_n}\ge \max_{s_n \le l \le e_n}{A_l} +  \sum_{l=s_n}^{e_n}{P_l}.
\end{equation}

\begin{figure*}[!t]
  \centering
  \includegraphics[width=1\textwidth]{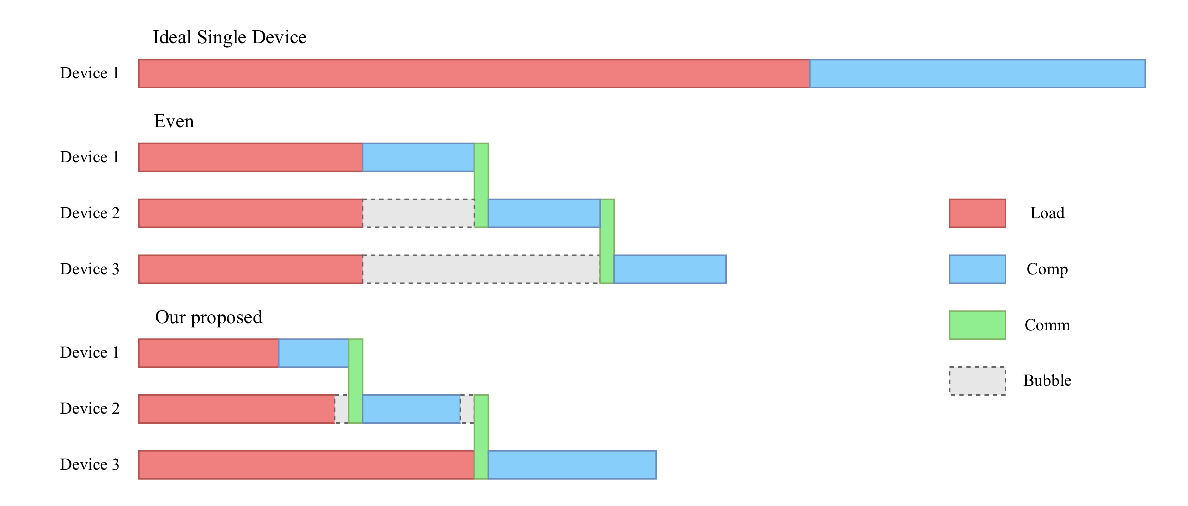}\\
  \caption{Gantt charts illustrating different allocation strategies on homogeneous devices, covering the entire cold start process including model loading, communication, and computation (prefill). There are two kinds of bubbles: obstructive and non-obstructive. The bubble after loading is an inevitable idle time while the bubble after computation can be filled by other requests. The former is what we aim to eliminate. In all strategy, the pipeline proceeds from top to bottom. \emph{Ideal Single Device} refers to assigning all model layers to the most powerful device in the system, assuming no RAM constraints. \emph{Even} denotes a uniform (even) distribution.}
  \label{fig:fig2}
\end{figure*}
\section{PROBLEM FORMULATION}\label{sec:III}

In this section, we first highlight the necessity of optimization by analyzing the performance bottlenecks of a baseline deployment strategy. We then formulate a combinatorial optimization problem aimed at minimizing the total cold-start latency:
\begin{equation}
    T = t^{\text{finish}}_{k_N}.
\end{equation}

To illustrate the core idea behind the proposed method, we first consider a baseline method in which the model is evenly partitioned into $K$ segments and deployed across $K$ homogeneous devices. As shown in Fig.~\ref{fig:fig2}, this strategy parallelizes model loading across devices but fails to exploit pipelining opportunities between model loading and computation. Specifically, each device $k_n$ must wait for its predecessor $k_{n-1}$ to complete both its computation and activation transfer before beginning its own computation. This results in idle time which can be quantified as:
\begin{equation}
    t^{\text{wait}}_{k_n} = \max(0, t^{\text{finish}}_{k_{n-1}} - t^{\text{load}}_{k_n}).
\end{equation}

To eliminate this bottleneck, the proposed algorithm aims to align the computation completion time of a predecessor device ($t^{\text{finish}}_{k_{n-1}}$) with the model loading time of its successor ($t^{\text{load}}_{k_n}$) through fine-grained layer allocation optimization. Ideally, this alignment ensures that $t^{\text{wait}}_{k_n} \approx 0$, enabling an efficient pipeline between the predecessor's computation and the successor's loading phase, thereby reducing end-to-end cold-start latency. This core principle extends naturally to more complex, heterogeneous environments. Even when devices differ in computational capacity ($c_k$) and communication resources ($r_k, b_k$), effective overlap between computation and loading can still be achieved by identifying an optimal, non-uniform model partitioning.

Based on the above analysis, we formalize the cold-start latency minimization problem as follows:
\begin{align*}
\text{\textbf{P0}:}\quad
    \min_{N,\mathbf{k,s,e}} \quad & T \\
    \text{s.t.} \quad
    &\mathcal{N}=\{1,2,...,N\}\tag{c1} \label{eq:constraint0}\\
    &n\neq n' \Rightarrow k_{n}\neq k_{n'},\quad \forall n, n'\in\mathcal{N} \label{eq:constraint1}\tag{c2} \\
    &s_1 = 1 \label{eq:constraint2}\tag{c3} \\
    &e_N = L \label{eq:constraint3}\tag{c4} \\
    &s_n \le e_n,\quad \forall n\in\mathcal{N} \label{eq:constraint4}\tag{c5} \\
    &s_{n} = e_{n-1} + 1,\quad \forall n\in\mathcal{N}\setminus\{1\} \label{eq:constraint5}\tag{c6} \\
    &m_{k_n}\ge \max_{s_n \le l \le e_n}{A_l} +  \sum_{l=s_n}^{e_n}{P_l},\quad \forall n \in\mathcal{N}\tag{c7}
\end{align*}
where $\mathbf{k}=(k_1,k_2,...,k_N)$, $\mathbf{s}=(s_1,s_2,...,s_N)$ and $\mathbf{e}=(e_1,e_2,...,e_N)$ denote the vectors of the assigned device, starting layer indices, and ending layer indices, respectively.

The optimization problem (P0) is a complex Mixed-Integer Non-Linear Program (MINLP). Its complexity arises primarily from two factors: (1) the presence of integer decision variables, including the number of partitions $N$, the layer boundaries $s_n, e_n$, and the combinatorial device assignments $k_n$; and (2) the nonlinearities introduced by the use of $\max$ and $\min$ operators in the timing constraints. Overall, \textbf{P0} is an NP-hard problem.

Despite its theoretical intractability, we observe that in practical deployment scenarios, the problem size is often moderate, e.g., the number of available devices $K$ and model layers $L$ typically satisfy $K \le 20, L \le 100$). More importantly, the problem exhibits a clear optimal substructure property, meaning that an optimal solution to the full problem can be built from optimal solutions to its subproblems. This property makes the problem well-suited for a DP approach. Accordingly, we design a DP algorithm capable of computing the exact optimal solution within a reasonable runtime, as detailed in the following section.

\section{PROPOSED DP ALGORITHM}\label{sec:IV}
\begin{algorithm}
\caption{Optimal Pipeline Scheduling}
\label{alg:optimal_pipeline}
\begin{algorithmic}[1]
\REQUIRE Number of devices $K$, number of model layers $L$, device parameters $\{\psi_1,\psi_2,...,\psi_K\}$, model parameters $\{\phi_1,\phi_2,...,\phi_L\}$
\ENSURE Minimum cold-start latency $T$, allocation plan $plan$
\STATE // Preprocessing
\STATE Compute and store all $M(i,j)$, $T_{\text{load}}(i,j,d)$, $T_{\text{comp}}(i,j,d)$, and $T_{\text{comm}}(d',d,i)$ values.
\STATE Initialize $DP(S,j,d) \gets \infty$ for all states.
\STATE Initialize $path(S,j,d) \gets \text{null}$ for backtracking.
\STATE // Base Cases
\STATE $DP(S,j,d) \gets T_{\text{load}}(1,j,d) + T_{\text{comp}}(1,j,d)$ for all $j\le L$, $d\le K$, $S=(1\ll (d-1))$.
\STATE $path(S,j,d) \gets (0,0)$ for all $j\le L$, $d\le K$, $S=(1\ll (d-1))$.
\STATE // DP Iterations
\FOR{$j \gets 2$ to $L$}
    \FOR{$S \gets 1$ to $((1 \ll K)-1)$}
        \IF{$\text{population\_count}(S) > 1$}
            \FOR{$d \gets $ to $K$}
                \IF{$((S \gg (d-1)) \,\&\, 1) = 1$}
                    \STATE $S' \gets S \oplus (1 \ll (d-1))$
                    \FOR{$i \gets 1$ to $j-1$}
                        \IF{$M(i,j) \le m_d$}
                            \FOR{$d' \gets 1$ to $K$}
                                \IF{$((S' \gg (d'-1)) \,\&\, 1) = 1$}
                                    \STATE $t_{\text{prev}} \gets DP(S',i,d')$
                                    \IF{$t_{\text{prev}} < \infty$}
                                        \STATE $t_{\text{finish}} \gets \max(T_{\text{load}}(i+1,j,d), t_{\text{prev}}) + T_{\text{comm}}(d',d,i) + T_{\text{comp}}(i+1,j,d)$
                                        \IF{$t_{\text{finish}} < DP(S,L,d)$}
                                            \STATE $DP(S,j,d) \gets t_{\text{finish}}$
                                            \STATE $path(S,j,d) \gets (i,d')$
                                        \ENDIF
                                    \ENDIF
                                \ENDIF
                            \ENDFOR
                        \ENDIF
                    \ENDFOR
                \ENDIF
            \ENDFOR
        \ENDIF
    \ENDFOR
\ENDFOR
\STATE Find $(S^*, d^*) = \arg\min_{S,d} \{DP(S,L,d)\}$
\STATE $T\gets DP(S^*,L,d^*)$
\STATE Reconstruct $plan$ by backtracking from $path(S^*,L,d^*)$.
\RETURN $T$, $plan$
\end{algorithmic}
\end{algorithm}

In the previous section, we formulated the problem as a MINLP, which involves jointly determining the model partitioning and computational task assignment. Due to the presence of nonlinear and non-convex operators, the problem cannot be efficiently solved using brute-force methods, which require exhaustive search with a computational complexity of $O((KL)^K)$. This level of complexity makes the approache infeasible for practical scenarios. Fortunately, the problem exhibits two key properties that make it well-suited for a DP solution:
\begin{itemize}
    \item \textbf{Optimal Substructure: }A globally optimal solution for partitioning and scheduling $L$ layers across $K$ devices necessarily includes optimal solutions to its subproblems, such as scheduling the first $j$ layers (where $j<L$) across a subset of devices $S\subset K$.
    \item \textbf{Overlapping Subproblems: } In the recursive construction of the global solution, certain subproblems, e.g., computing the minimum completion time for scheduling, the first $j$ layers on a given subset of devices recur multiple times. DP mitigates redundant computations by storing and reusing these intermediate results.
\end{itemize}

\subsection{DP State Definition}
We begin by defining the following utility functions for time cost and memory usage, which can be precomputed prior to executing the main DP algorithm to enhance efficiency:
\begin{align}
    T_{\text{load}}(i,j,d)&=\frac{\sum_{l=i}^j{P_l}}{r_d},\\
    T_{\text{comp}}(i,j,d)&=\frac{\sum_{l=i}^{j} W_l}{c_d},\\
    T_{\text{comm}}(d_{\text{prev}},d_{\text{curr}},j)&=\frac{A_j}{\min{(b^{u}_{d_{\text{prev}}},b^{d}_{d_{\text{curr}}})}},\\
    M(i,j) &= \max_{i \le l \le j}{A_l}+\sum_{l=i}^j{P_l}.
\end{align}

To formulate the subproblems and define the state transitions in our DP approach, we represent each state as a triplet $(S,j,d)$, where $S\subset K$ is the set of devices used so far, $j\in \{1,2,...,L\}$ is the index of the last layer assigned. $d\in S$ is the device to which the final segment (ending at layer $j$) is assigned. In our implementation, the set $S$ is represented by a bitmask, an integer where the $k$-th (binary) bit is 1 if device $d$ is in the set, and 0 otherwise.

\begin{table*}[!t]
\setlength{\tabcolsep}{2pt}
\belowrulesep=0pt 
\aboverulesep=0pt 
\renewcommand{\arraystretch}{1.4} 
\doublerulesep 2.2pt
\caption{Simulation Parameters.}
\label{tab:tab1}
\footnotesize
\begin{tabular*}{\textwidth}{@{\extracolsep{\fill}}c|cccccccccccccr}
\toprule
    \multirow{2}{*}{{\centering Parameters}}& \multicolumn{3}{c}{Computation}&\multicolumn{2}{c}{Load}&\multicolumn{9}{c}{Communication}\\[-0.2em]
    \cmidrule(lr){2-4}\cmidrule(lr){5-6}\cmidrule(l){7-15} 
    &$c^*$ (FLOPS) & $a$ & $b$ & $r_k$ (MB/s) & $m_k$ (GB) & $\mu$ & $B_k$ (MHz) & $P^u_k$ (dBm) & $P^d_k$ (dBm) & $N_0$ (dBm/Hz) & $d_k$ (m) & $d_0$ (m) & $\zeta$ & $\beta_0$ (dB) \\ \hline
    Device 1 & 165 & 0.4 & 5.1E-04 & 5000 & 20 & \multirow{4}{*}{0.5} & \multirow{4}{*}{160} & 20 & \multirow{4}{*}{25} & \multirow{4}{*}{-174} & 1 & \multirow{4}{*}{1} & \multirow{4}{*}{3} & \multirow{4}{*}{-47.2} \\
    Device 2 & 70 & 0.7 & 8.7E-04 & 4000 & 10 &  &  & 18 &  &  & 3 &  &  & \\
    Device 3 & 30 & 0.8 & 1.1E-03 & 3000 & 8 &  &  & 15 &  &  & 5 &  &  & \\
    Device 4 & 20 & 0.8 & 1.8E-03 & 2000 & 8 &  &  & 15 &  &  & 7 &  &  & \\
\bottomrule
\end{tabular*}
\end{table*}

The DP state value, denoted as $DP(S, j, d)$, represents the minimum total completion time required to partition and schedule the first $j$ layers (i.e., layers 1 to $j$) onto the device set $S$, with the final segment processed by device $d$.
\subsection{State Transition Recurrence}
To compute the value of the DP state $DP(S,j,d)$, we consider all possible predecessor states corresponding to valid partitions of the first $j$ layers. Suppose the final segment assigned to device $d$ begins at layer $i+1$, where $1\le i<j$). This implies that the first $i$ layers (i.e., layer $1$ through $i$) have already been scheduled on the reduced device set $S'=S\setminus\{d\}$, with the final segment in that subproblem executed on some device $d'\in S'$. The state transition recurrence can be written as:
\begin{multline}
DP(S,j,d) = \min_{\substack{1 \le i < j, \\ d' \in S'}} \left\{ 
\max\left(T_{\text{load}}(i+1, j, d), DP(S',i,d')\right) \right. \\
\left. + T_{\text{comm}}(d', d, i) + T_{\text{comp}}(i+1, j, d) 
\right\}.
\end{multline}

This recurrence captures the dependencies between partition boundaries, device assignments, and timing, ensuring that model loading on device $d$ and computation on the previous device $d'$ are appropriately overlapped while respecting communication delays.

\subsection{Base Case}
The base case of the recurrence corresponds to the scenario where the entire pipeline consists of a single segment, where all layers from $1$ to $j$ are assigned to a single device $d$. In this case, there is no predecessor device, and therefore no inter-device communication. The set of used devices is $S=\{d\}$. The corresponding base case is defined as:
\begin{multline}
    DP(1\ll (d-1),j,d)=T_{load}(1,j,d)+T_{comp}(1,j,d),\\
    \forall d\in\mathcal{K},j\in\mathcal{L},\quad M(1,j)\le m_d.
\end{multline}

\subsection{Reconstructing the Optimal Schedule}
To recover the optimal scheduling plan after computing the DP table, we store, for each state $DP(S,L,d)$, the choice of split point $i$ and predecessor device $d'$ that led to the minimum value. Once the DP table is fully populated, the optimal total makespan is given by
\begin{equation}
    T = \min_{S\subset\mathcal{K},S\neq \emptyset}{\left(\min_{d\in S}(DP(S,L,d))\right)}.
\end{equation}

To enable reconstruction of the complete solution, we maintain a backtracking table path $path(S,j,d)=(i,d')$, where $i$ is the starting layer of the final segment assigned to device $d'$. By recursively following these stored paths starting from the final state, we can reconstruct the entire layer-to-device assignment and scheduling order.

The full procedure is formally described in the accompanying pseudocode. The DP iteration involves five nested loops, resulting in a space complexity of $O(K\cdot L\cdot 2^K)$ and a time complexity of $O(K^2\cdot L^2\cdot 2^K)$. While the algorithm remains exponential in $K$, this complexity is significantly more tractable than the factorial complexity of brute-force enumeration, and is acceptable for practical values of $K$ and $L$.

\begin{figure*}[!t]
    \captionsetup[subfloat]{captionskip=-4pt} 
    \captionsetup[subfloat]{nearskip=-6pt}
    \centering
    \subfloat[]{\includegraphics[width=0.45\textwidth]{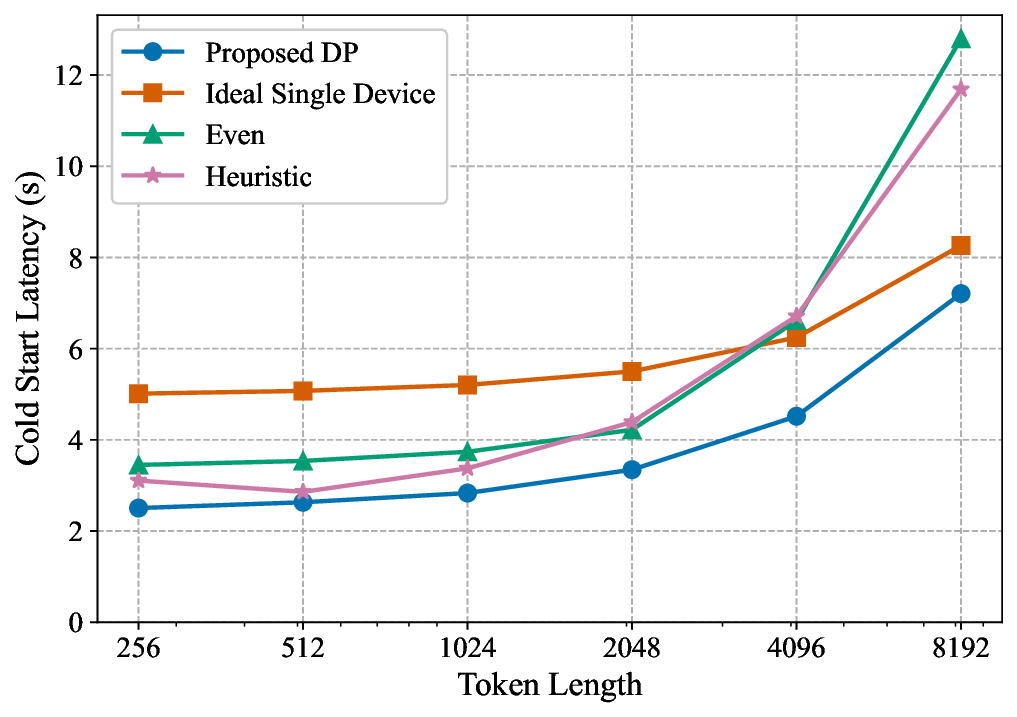}\label{fig:fig3a}}
    \hfill
    \subfloat[]{\includegraphics[width=0.45\textwidth]{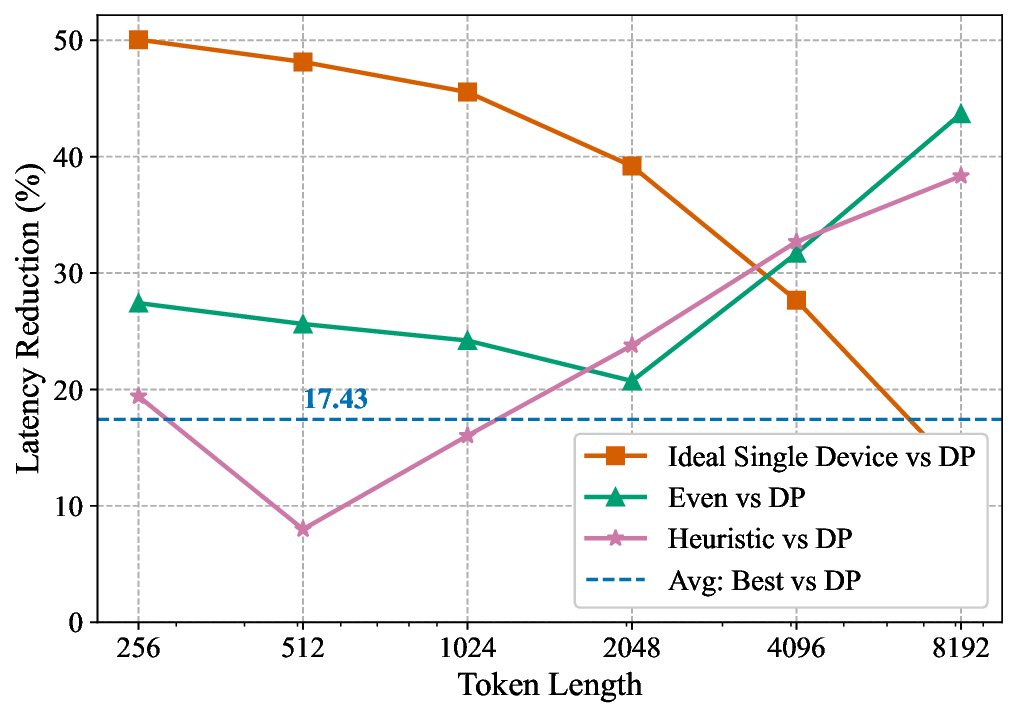}\label{fig:fig3b}}
    \caption{Comparison of four strategies. (a) Cold start delay changes with load; (b) Performance improvement percentage of our algorithm compared to baseline strategies.}
    \label{fig:fig3}
\end{figure*}

\begin{figure*}[!t]
\captionsetup[subfloat]{captionskip=-2pt,nearskip=-6pt} 
\centering
\subfloat[]{\includegraphics[width=0.45\textwidth]{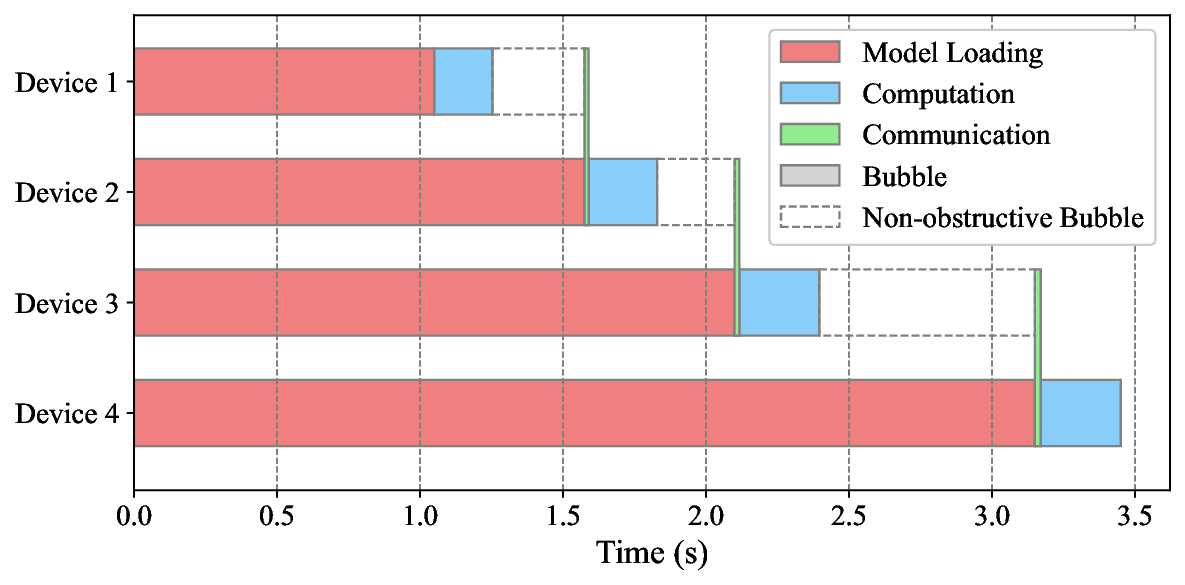}\label{fig:fig4a}}
\hfill
\subfloat[]{\includegraphics[width=0.45\textwidth]{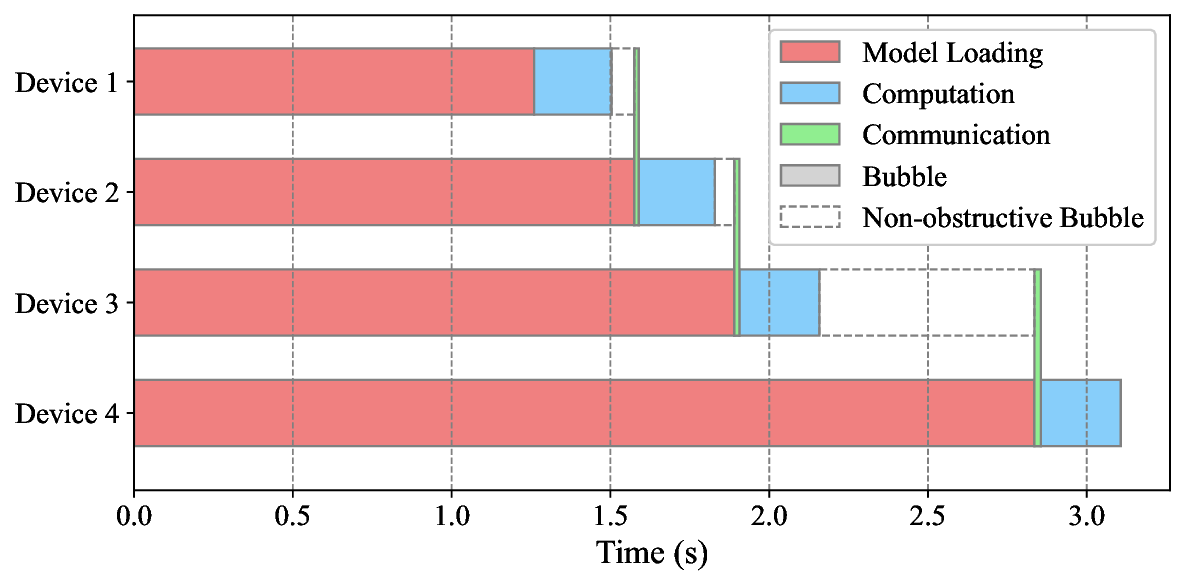}\label{fig:fig4b}}\\
\subfloat[]{\includegraphics[width=0.45\textwidth]{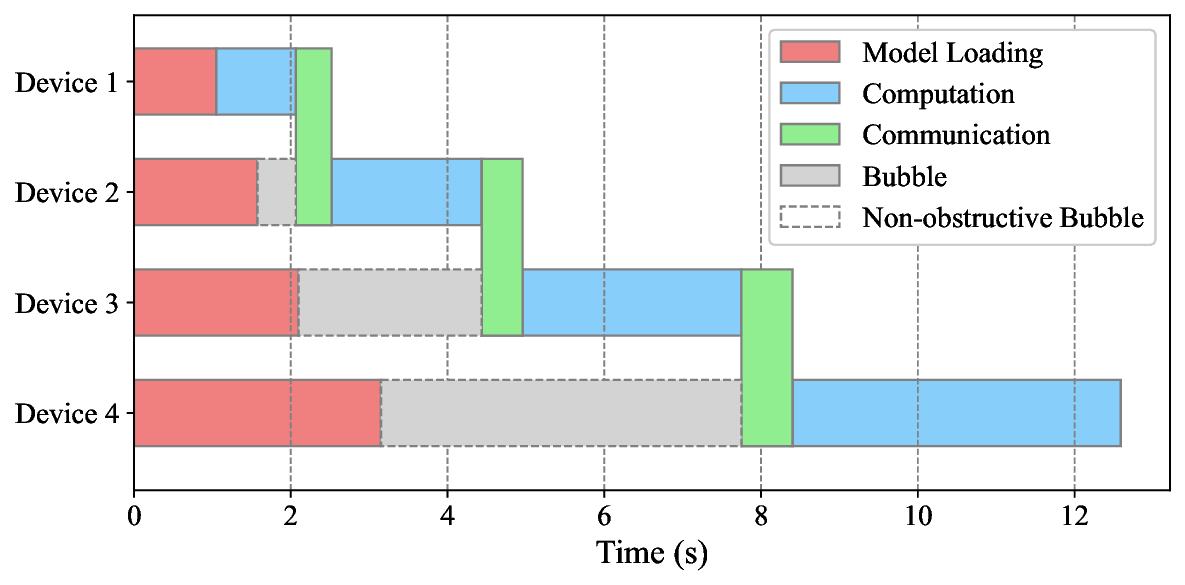}\label{fig:fig4c}}
\hfill
\subfloat[]{\includegraphics[width=0.45\textwidth]{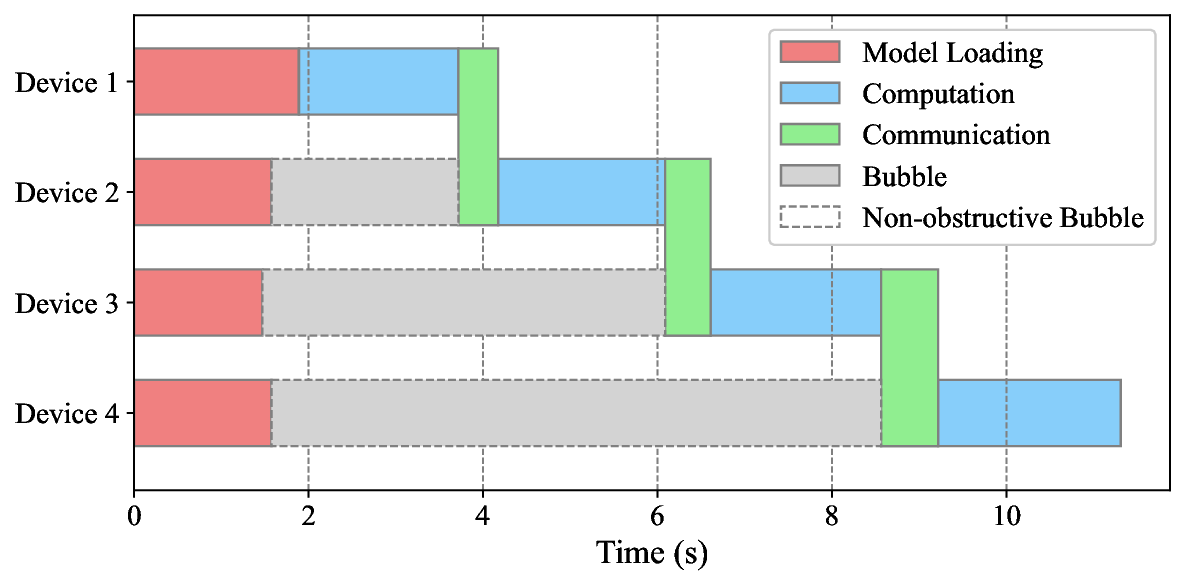}\label{fig:fig4d}}
\caption{Gantt chart of Even and Heuristic strategies. (a) Even strategy for 256 token length; (b) Even strategy for 8196 token length; (c) Heuristic strategy for 256 token length; (d) Heuristic strategy for 8196 token length.}
\label{fig:fig4}
\end{figure*}

\section{NUMERICAL RESULTS}\label{sec:V}
\subsection{Experiment Setting}
To evaluate the performance of the proposed algorithm, we conduct a series of numerical simulations. The simulation environment is configured as a typical personal Wi-Fi network, consisting of one AP and four heterogeneous computing devices. For model parameters, we adopt the settings of the Qwen3-14B\cite{Qwen} model. The detailed simulation configurations, including the computational, I/O characteristics, and communication bandwidths of each device, are summarized in Tab~\ref{tab:tab1}.

\subsection{Performance Evaluation}
To comprehensively validate the effectiveness of the proposed algorithm, we compare it against three baseline strategies:
\begin{itemize}
    \item \textbf{Ideal Single Device:} An idealized reference in which all model layers are deployed on the most powerful device (Device 1), assumed to have unlimited GPU memory.
    \item \textbf{Even:} A naive strategy that evenly partitions model layers among the four devices, ignoring performance heterogeneity. Allocation priority is given to stronger devices.
    \item \textbf{Heuristic:} A performance-aware strategy that allocates layers based on a metric defined as the harmonic mean of a device’s compute capability and disk read speed. Layers are proportionally distributed according to this metric, again prioritizing stronger devices.
\end{itemize}

As shown in Fig.\ref{fig:fig3}, the proposed DP algorithm consistently achieves the lowest cold-start latency across all evaluated token lengths (256 to 8192), significantly outperforming all baseline methods. Fig.\ref{fig:fig3b} further indicates that our method reduces latency by $8\%$ to $50\%$ compared to the baselines. Even against the best-performing baseline at each token length, it delivers an average improvement of $17.43\%$. The following analysis reveals a key trade-off underlying these results: I/O parallelization versus computational heterogeneity.

\begin{figure*}[!t]
\captionsetup[subfloat]{captionskip=0pt,nearskip=-6pt} 
\centering
\subfloat[]{\includegraphics[width=0.33\textwidth]{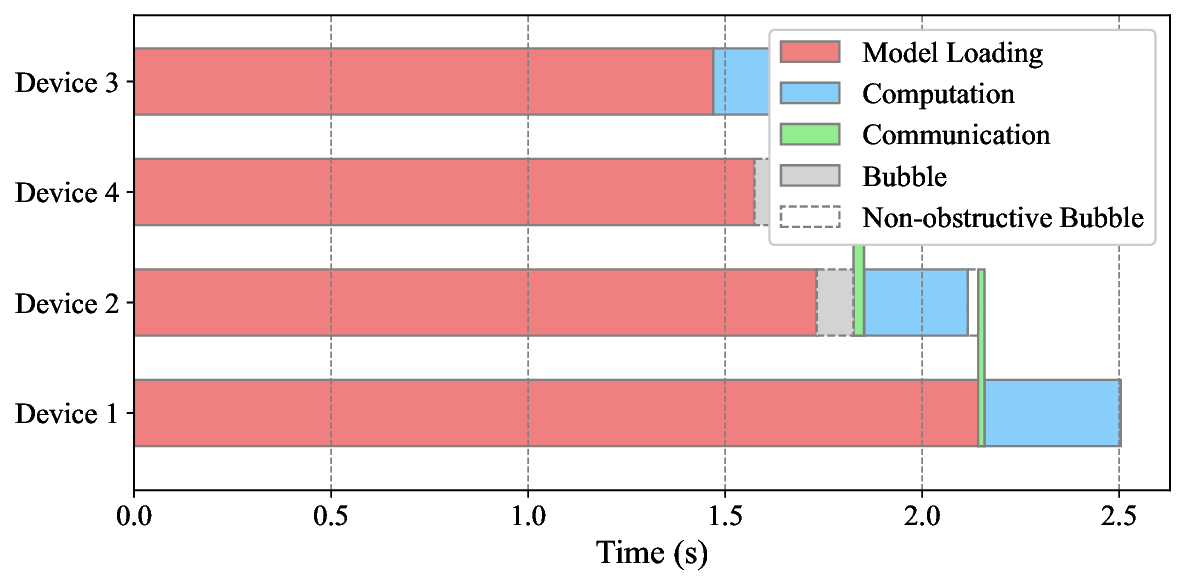}\label{fig:fig5a}}
\hfill
\subfloat[]{\includegraphics[width=0.33\textwidth]{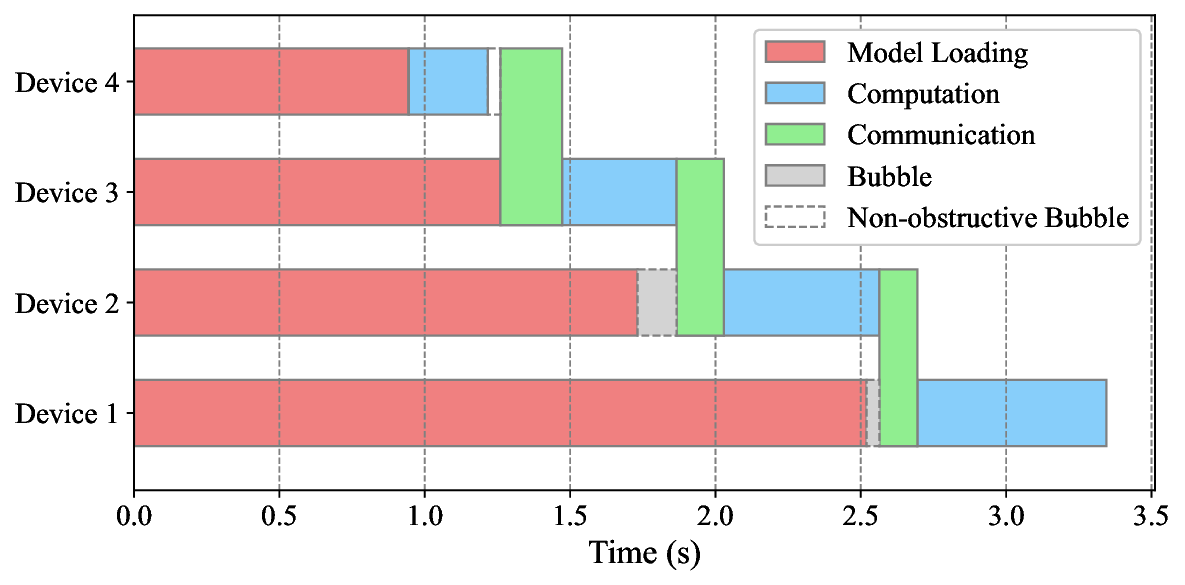}\label{fig:fig5b}}
\hfill
\subfloat[]{\includegraphics[width=0.33\textwidth]{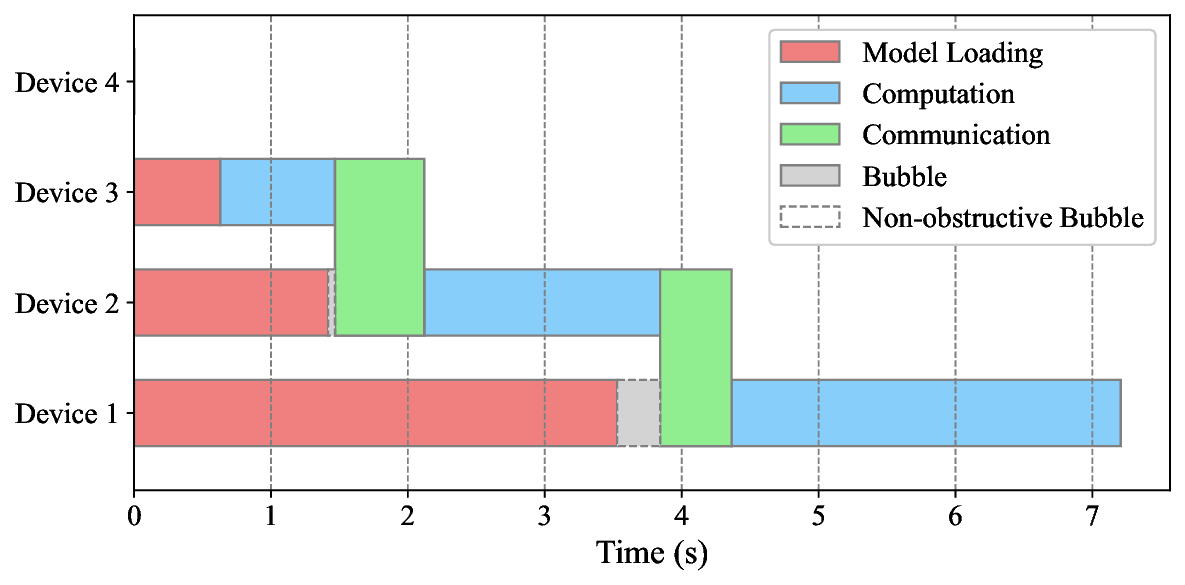}\label{fig:fig5c}}
\hfill
\caption{Gantt chart of our proposed DP algorithm. (a) 256 token length; (b) 2048 token length; (c) 8196 token length.}
\label{fig:fig5}
\end{figure*}

1) \textbf{Limitations of Baseline Strategies:} In short-token scenarios (e.g., $\le$ 2048 tokens), the \emph{Even} and \emph{Heuristic} strategies perform reasonably well by parallelizing model loading during the I/O-intensive phase, as illustrated in Fig.~\ref{fig:fig4a} and Fig.~\ref{fig:fig4c}. However, as token length increases, computation becomes the dominant bottleneck, revealing the limitations of these static allocation strategies as depicted in Fig.~\ref{fig:fig4b} and Fig.~\ref{fig:fig4d}. Because they fail to consider computational heterogeneity, the weakest device often becomes a pipeline ``straggler'', significantly increasing overall latency. In contrast, the Ideal Single Device exhibits the opposite behavior: although it suffers from higher latency in the I/O phase due to lack of parallelism, its strong computational power enables it to outperform the imbalanced Even and Heuristic strategies in compute-intensive phases.

2) \textbf{Adaptive Advantage of the Proposed DP:} The Proposed DP algorithm effectively addresses the trade-off between I/O and computation. Rather than using a fixed allocation scheme, it dynamically optimizes layer deployment based on the input token length, which determines the expected computational workload. Under low-load conditions, the algorithm favors a more balanced layer distribution to maximize I/O parallelism, as shown in Fig.~\ref{fig:fig5a}. Conversely, under high-load conditions, it strategically assigns more compute-intensive layers to the most powerful devices, preventing weaker devices from becoming bottlenecks and fully leveraging the system’s heterogeneous computing resources, as illustrated in Fig.\ref{fig:fig5c}. This shift in allocation strategy with increasing load is clearly visible in Fig.~\ref{fig:fig5}: as the load grows, the distribution becomes more skewed, with a larger proportion of layers assigned to stronger devices.

In summary, by accurately modeling and dynamically optimizing computation, communication, and I/O overheads, the proposed DP algorithm effectively adapts to device heterogeneity, enabling it to achieve near-optimal performance across diverse workload conditions.

\section{CONCLUSION AND FUTURE WORK}
In this paper, we tackle the challenge of inference cold-start latency in wireless networks composed of heterogeneous devices. We propose a novel dynamic programming–based layer allocation algorithm to minimize latency. By carefully modeling computation, communication, and I/O overheads, our approach effectively balances these factors to fully leverage the parallel processing capabilities of multiple devices. Experimental results demonstrate that our algorithm significantly outperforms several baseline strategies in reducing end-to-end latency.

Looking forward, a key direction for future work is the seamless integration of our cold-start optimization with strategies tailored for the steady-state, high-throughput inference phase. Designing a dynamic and adaptive transition mechanism between these phases is a critical challenge for achieving comprehensive, lifecycle-aware latency minimization. Our future research will focus on developing a unified optimization framework to address this challenge.


\bibliographystyle{gbt7714-numerical}
\bibliography{bib}

\begin{thebibliography}{29}
\providecommand{\natexlab}[1]{#1}
\providecommand{\url}[1]{#1}
\expandafter\ifx\csname urlstyle\endcsname\relax\else
  \urlstyle{same}\fi
\expandafter\ifx\csname href\endcsname\relax
  \DeclareUrlCommand\doi{\urlstyle{rm}}
  \def\eprint#1#2{#2}
\else
  \def\doi#1{\href{https://doi.org/#1}{\nolinkurl{#1}}}
  \let\eprint\href
\fi

\bibitem[Yang et~al.(2023)Yang, Zhang, Song, Hong, Xu, Zhao, Zhang, Cui, and
  Yang]{diffusion}
YANG L, ZHANG Z, SONG Y, et~al.
\newblock Diffusion models: A comprehensive survey of methods and
  applications\allowbreak[J].
\newblock ACM computing surveys, 2023, 56\allowbreak (4): 1-39.

\bibitem[Liu et~al.(2024)Liu, Huang, Li, Wang, and Xiao]{privacy}
LIU Y, HUANG J, LI Y, et~al.
\newblock Generative ai model privacy: a survey\allowbreak[J].
\newblock Artificial Intelligence Review, 2024, 58\allowbreak (1): 33.

\bibitem[Rodriguez-Conde et~al.(2023)Rodriguez-Conde, Campos, and
  Fdez-Riverola]{inference}
RODRIGUEZ-CONDE I, CAMPOS C, FDEZ-RIVEROLA F.
\newblock Horizontally distributed inference of deep neural networks for
  ai-enabled iot\allowbreak[J].
\newblock Sensors, 2023, 23\allowbreak (4): 1911.

\bibitem[Liu et~al.(2025)Liu, Tao, Cao, and Ming]{parallelization}
LIU S, TAO X, CAO W, et~al.
\newblock Parallelization techniques for large language models: A review from
  training to inference\allowbreak[C]//\allowbreak
International Conference on Wireless Artificial Intelligent Computing Systems
  and Applications.
\newblock Springer, 2025: 307-317.

\bibitem[Huang et~al.(2019)Huang, Cheng, Bapna, Firat, Chen, Chen, Lee, Ngiam,
  Le, Wu, et~al.]{gpipe}
HUANG Y, CHENG Y, BAPNA A, et~al.
\newblock Gpipe: Efficient training of giant neural networks using pipeline
  parallelism\allowbreak[J].
\newblock Advances in neural information processing systems, 2019, 32.

\bibitem[Narayanan et~al.(2019)Narayanan, Harlap, Phanishayee, Seshadri,
  Devanur, Ganger, Gibbons, and Zaharia]{pipedream}
NARAYANAN D, HARLAP A, PHANISHAYEE A, et~al.
\newblock Pipedream: Generalized pipeline parallelism for dnn
  training\allowbreak[C]//\allowbreak
Proceedings of the 27th ACM symposium on operating systems principles.
\newblock 2019: 1-15.

\bibitem[Jeon et~al.(2025)Jeon, Wu, Cao, Kim, Park, Aggarwal, Unger, Arfeen,
  Liao, Miao, et~al.]{graphpipe}
JEON B, WU M, CAO S, et~al.
\newblock Graphpipe: Improving performance and scalability of dnn training with
  graph pipeline parallelism\allowbreak[C]//\allowbreak
Proceedings of the 30th ACM International Conference on Architectural Support
  for Programming Languages and Operating Systems, Volume 1.
\newblock 2025: 557-571.

\bibitem[Qi et~al.(2023)Qi, Wan, Huang, and Lin]{zero}
QI P, WAN X, HUANG G, et~al.
\newblock Zero bubble pipeline parallelism\allowbreak[A].
\newblock 2023.

\bibitem[Fang et~al.(2024)Fang, Pan, Wang, Li, and Sun]{pipefusion}
FANG J, PAN J, WANG J, et~al.
\newblock Pipefusion: Patch-level pipeline parallelism for diffusion
  transformers inference\allowbreak[A].
\newblock 2024.

\bibitem[Hu et~al.(2022)Hu, Imes, Zhao, Kundu, Beerel, Crago, and
  Walters]{pipeedge}
HU Y, IMES C, ZHAO X, et~al.
\newblock Pipeedge: Pipeline parallelism for large-scale model inference on
  heterogeneous edge devices\allowbreak[C]//\allowbreak
2022 25th Euromicro Conference on Digital System Design (DSD).
\newblock IEEE, 2022: 298-307.

\bibitem[Yoon et~al.(2021)Yoon, Byeon, Kim, and Lee]{edgepipe}
YOON J, BYEON Y, KIM J, et~al.
\newblock Edgepipe: Tailoring pipeline parallelism with deep neural networks
  for volatile wireless edge devices\allowbreak[J].
\newblock IEEE Internet of Things Journal, 2021, 9\allowbreak (14):
  11633-11647.

\bibitem[Xiong et~al.(2024)Xiong, Liu, Zhang, Zu, Zhu, and Zhou]{epipe}
XIONG Y, LIU W, ZHANG R, et~al.
\newblock Epipe: Pipeline inference framework with high-quality offline
  parallelism planning for heterogeneous edge
  devices\allowbreak[C]//\allowbreak
Proceedings of the 43rd IEEE/ACM International Conference on Computer-Aided
  Design.
\newblock 2024: 1-10.

\bibitem[Xue et~al.(2024)Xue, Sun, Chen, Tao, Xu, Qian, Cui, and Zhang]{wdmoe}
XUE N, SUN Y, CHEN Z, et~al.
\newblock Wdmoe: Wireless distributed large language models with mixture of
  experts\allowbreak[C]//\allowbreak
GLOBECOM 2024-2024 IEEE Global Communications Conference.
\newblock IEEE, 2024: 2707-2712.

\bibitem[Yao et~al.(2021)Yao, Dong, Zheng, Gholami, Yu, Tan, Wang, Huang, Wang,
  Mahoney, et~al.]{hawq}
YAO Z, DONG Z, ZHENG Z, et~al.
\newblock Hawq-v3: Dyadic neural network
  quantization\allowbreak[C]//\allowbreak
International Conference on Machine Learning.
\newblock PMLR, 2021: 11875-11886.

\bibitem[Lin et~al.(2024)Lin, Tang, Tang, Yang, Chen, Wang, Xiao, Dang, Gan,
  and Han]{awq}
LIN J, TANG J, TANG H, et~al.
\newblock Awq: Activation-aware weight quantization for on-device llm
  compression and acceleration\allowbreak[J].
\newblock Proceedings of machine learning and systems, 2024, 6: 87-100.

\bibitem[Fang et~al.(2023)Fang, Ma, Song, Mi, and Wang]{dep}
FANG G, MA X, SONG M, et~al.
\newblock Depgraph: Towards any structural pruning\allowbreak[C]//\allowbreak
Proceedings of the IEEE/CVF conference on computer vision and pattern
  recognition.
\newblock 2023: 16091-16101.

\bibitem[Gan et~al.(2022)Gan, Chen, Li, Chen, Cheng, Wang, Liu, Wang, and
  Liu]{playing}
GAN Z, CHEN Y~C, LI L, et~al.
\newblock Playing lottery tickets with vision and
  language\allowbreak[C]//\allowbreak
Proceedings of the AAAI Conference on Artificial Intelligence: Vol.~36.
\newblock 2022: 652-660.

\bibitem[Gu et~al.(2023)Gu, Dong, Wei, and Huang]{minillm}
GU Y, DONG L, WEI F, et~al.
\newblock Minillm: Knowledge distillation of large language
  models\allowbreak[A].
\newblock 2023.

\bibitem[Sauer et~al.(2024)Sauer, Lorenz, Blattmann, and Rombach]{adversarial}
SAUER A, LORENZ D, BLATTMANN A, et~al.
\newblock Adversarial diffusion distillation\allowbreak[C]//\allowbreak
European Conference on Computer Vision.
\newblock Springer, 2024: 87-103.

\bibitem[Vaswani et~al.(2017)Vaswani, Shazeer, Parmar, Uszkoreit, Jones, Gomez,
  Kaiser, and Polosukhin]{transformer}
VASWANI A, SHAZEER N, PARMAR N, et~al.
\newblock Attention is all you need\allowbreak[J].
\newblock Advances in neural information processing systems, 2017, 30.

\bibitem[Ainslie et~al.(2023)Ainslie, Lee-Thorp, de~Jong, Zemlyanskiy, Lebron,
  and Sanghai]{GQA}
AINSLIE J, LEE-THORP J, DE~JONG M, et~al.
\newblock Gqa: Training generalized multi-query transformer models from
  multi-head checkpoints\allowbreak[C]//\allowbreak
Proceedings of the 2023 Conference on Empirical Methods in Natural Language
  Processing.
\newblock 2023: 4895-4901.

\bibitem[Shazeer(2020)]{GLU}
SHAZEER N.
\newblock Glu variants improve transformer\allowbreak[A].
\newblock 2020.

\bibitem[Kirk et~al.(2016)Kirk and Wen-Mei]{programming}
KIRK D~B, WEN-MEI W~H.
\newblock Programming massively parallel processors: a hands-on
  approach\allowbreak[M].
\newblock Morgan kaufmann, 2016.

\bibitem[Hennessy et~al.(2011)Hennessy and Patterson]{computer}
HENNESSY J~L, PATTERSON D~A.
\newblock Computer architecture: a quantitative approach\allowbreak[M].
\newblock Elsevier, 2011.

\bibitem[Lanante et~al.(2017)Lanante, Uwai, Nagao, Kurosaki, and
  Ghosh]{performance}
LANANTE L, UWAI H~O~T, NAGAO Y, et~al.
\newblock Performance analysis of the 802.11 ax ul ofdma random access protocol
  in dense networks\allowbreak[C]//\allowbreak
2017 IEEE international conference on communications (ICC).
\newblock IEEE, 2017: 1-6.

\bibitem[Qu et~al.(2019)Qu, Li, Yang, Yan, Yang, Deng, and Chen]{survey}
QU Q, LI B, YANG M, et~al.
\newblock Survey and performance evaluation of the upcoming next generation
  wlans standard-ieee 802.11 ax\allowbreak[J].
\newblock Mobile Networks and Applications, 2019, 24\allowbreak (5): 1461-1474.

\bibitem[Shannon(1948)]{shannon}
SHANNON C~E.
\newblock A mathematical theory of communication\allowbreak[J].
\newblock The Bell system technical journal, 1948, 27\allowbreak (3): 379-423.

\bibitem[Rappaport(2010)]{wireless}
RAPPAPORT T~S.
\newblock Wireless communications: Principles and practice, 2/e\allowbreak[M].
\newblock Pearson Education India, 2010.

\bibitem[Yang et~al.(2025)Yang, Li, Yang, Zhang, Hui, Zheng, Yu, Gao, Huang,
  Lv, et~al.]{Qwen}
YANG A, LI A, YANG B, et~al.
\newblock Qwen3 technical report\allowbreak[A].
\newblock 2025.

\end{thebibliography}














\section*{About the Authors}\footnotesize\vskip 2mm

\noindent{\bf Xuran Liu} was born in Anhui Province, China in 2003. He is currently a senior undergraduate student at the Shanghai Jiao Tong University, Shanghai, majoring in Information Engineering. His research interests focus on the efficient inference and deployment of large-scale models at the wireless edge.
\vskip4.59mm

\noindent{\bf Nan Xue} [corresponding author] received the B.E. degree in Telecommunication Engineering from Xidian University, Xi’an, China, in 2023. He is currently pursuing the Ph.D degree with the Department of Information and Communication Engineering. His research interests include wireless networks for large language models.
\vskip4.59mm

\noindent{\bf Rui Bao} received his B.S. degree in Information Engineering from the IEEE Honor Class at Shanghai Jiao Tong University in 2025. He will pursue a Ph.D. in Information and Communication Engineering at SJTU under the supervision of Prof. Zhiyong Chen. His research interests include large language models, agents system, and wireless communication networks.
\vskip4.59mm

\noindent{\bf Yaping Sun} received the B.E. degree and the Ph.D. degree from Xidian University and Shanghai Jiao Tong University, in 2015 and 2020, respectively. From 2018 to 2019, she was a Visiting Scholar with University of Washington, USA. From 2020 to 2022, She was a Post-Doctoral Research Fellow with the Future Network of Intelligent Institute (FNii), the Chinese University of Hong Kong (CUHK), Shenzhen, China. She is currently an Assistant Researcher with the Department of Broadband, Peng Cheng Laboratory, Shenzhen, and also with the FNii-Shenzhen, the CUHK, Shenzhen, China. Her research interests include mobile 3C networks and semantic communication.
\vskip4.59mm

\noindent{\bf Zhiyong Chen} received the Ph.D. degree from the School of Information and Communication Engineering, Beijing University of Posts and Telecommunications (BUPT), Beijing, China, in 2011. From 2009 to 2011, he was a visiting Ph.D. Student at the Department of Electronic Engineering, University of Washington, Seattle, USA. He is currently a Professor with the Cooperative Medianet Innovation Center, Shanghai Jiao Tong University (SJTU), Shanghai, China. His research interests include mobile communications-computing-caching (3C) networks and mobile AI systems. He served as the Student Volunteer Chair for the IEEE ICC 2019, the Publicity Chair for the IEEE/CIC ICCC 2014 and a TPC member for major international conferences. He was the recipient of the IEEE Asia-Pacific Outstanding Paper Award in 2019.
\vskip4.59mm

\noindent{\bf Meixia Tao} is a Distinguished Professor with the Department of Electronic Engineering at Shanghai Jiao Tong University, China. She received the B.S. degree from Fudan University, Shanghai, China, in 1999, and the Ph.D. degree from the Hong Kong University of Science and Technology in 2003. Her current research interests include wireless edge learning, semantic communications, integrated communication-computing-sensing, AI-based chan- nel modeling and beamforming.

Dr. Tao receives the 2019 IEEE Marconi Prize Paper Award, the 2013 IEEE Heinrich Hertz Paper Award, and a number of IEEE conference best paper awards. She also receives the 2009 IEEE ComSoc Asia-Pacific Outstanding Young Researcher award. Dr. Tao is a member of IEEE ComSoc GLOBECOM/ICC Technical Content (GITC) Committee. She has served for IEEE TRANSACTIONS ON WIRELESS COMMUNICATIONS for many years, first as an associate editor, then as a member of the Executive Editorial Committee, and currently as a member of the Steering Committee. She was also on the editorial board as an editor or guest editor of a number of other journals, including the IEEE TRANSACTIONS ON OMMUNICATIONS, IEEE TRANSACTIONS ON INFORMATION THEORY, PROCEEDINGS OF THE IEEE, and IEEE JOURNAL ON SELECTED AREAS IN COMMUNICATIONS. She has served as the TPC Co-Chair of IEEE ICC 2023 and Symposia Oversight Chair of IEEE ICC 2019.
\vskip4.59mm

\noindent{\bf Xiaodong Xu} (S’06-M’07-SM’18) received his B.S degree and Master’s Degree both from Shandong University in 2001 and 2004 separately. He received his Ph.D. degree in Beijing University of Posts and Telecommunications (BUPT) in 2007. He is currently a professor of BUPT, a research fellow of the Department of Broadband Communication of Peng Cheng Laboratory and a member of IMT-2030 (6G) Experts Panel. He has coauthored nine books/chapters and more than 130 journal and conference papers. He is also the inventor or co-inventor of 73 granted patents. His research interests cover semantic communications, intellicise networks, trial system and network.
\vskip4.59mm

\noindent{\bf Shuguang Cui} received his Ph.D. from Stanford in 2005. He is now a X.Q. Deng Presidential Chair Professor at The Chinese University of Hong Kong, Shenzhen, China. His current research interest is data driven large-scale information analysis and system design. He was selected as the Thomson Reuters Highly Cited Researcher and listed in the Worlds’ Most Influential Scientific Minds by ScienceWatch in 2014. He was the recipient of the IEEE SP Society 2012 and ComSoc 2023 Marconi Best Paper Awards. He is an IEEE Fellow, Member of Both Royal Society of Canada and Canadian Academy of Engineering.
\end{document}